\documentclass[twocolumn]{aastex6}
\usepackage{graphicx}
\usepackage[flushleft]{threeparttable}

\shorttitle{ASPECS: CO luminosity functions and cosmic density of molecular gas}
\shortauthors{Decarli et al.}

\def\Msun{M$_\odot$}

\def\Cii{[C\,{\sc ii}]}

\def\kms{km\,s$^{-1}$}

\def\Kkmspc{K~km\,s$^{-1}$\,pc$^2$}

\def\lsim{\mathrel{\rlap{\lower 3pt \hbox{$\sim$}} \raise 2.0pt \hbox{$<$}}}
\def\gsim{\mathrel{\rlap{\lower 3pt \hbox{$\sim$}} \raise 2.0pt \hbox{$>$}}}

\begin{document}

\title{
ALMA spectroscopic survey in the Hubble Ultra Deep Field: CO luminosity functions and the evolution of the cosmic density of molecular gas}

\author{
Roberto Decarli\altaffilmark{1}, 
Fabian Walter\altaffilmark{1,2,3}, 
Manuel Aravena\altaffilmark{4}, 
Chris Carilli\altaffilmark{3,5}, 
Rychard Bouwens\altaffilmark{6}, 
Elisabete da Cunha\altaffilmark{7,8}, 
Emanuele Daddi\altaffilmark{9}, 
R.\,J.~Ivison\altaffilmark{10,11}, 
Gerg\"{o} Popping\altaffilmark{10}, 
Dominik Riechers\altaffilmark{12}, 
Ian Smail\altaffilmark{13}, 
Mark Swinbank\altaffilmark{14}, 
Axel Weiss\altaffilmark{15}, 
Timo Anguita\altaffilmark{15,16}, 
Roberto Assef\altaffilmark{4}, 
Franz Bauer\altaffilmark{17,18,19}, 
Eric F.~Bell\altaffilmark{20}, 
Frank Bertoldi\altaffilmark{21}, 
Scott Chapman\altaffilmark{22}, 
Luis Colina\altaffilmark{23}, 
Paulo C.~Cortes\altaffilmark{24,25}, 
Pierre Cox\altaffilmark{24}, 
Mark Dickinson\altaffilmark{26}, 
David Elbaz\altaffilmark{9}, 
Jorge G\'onzalez-L\'opez\altaffilmark{27}, 
Edo Ibar\altaffilmark{28}, 
Leopoldo Infante\altaffilmark{27}, 
Jacqueline Hodge\altaffilmark{6}, 
Alex Karim\altaffilmark{21}, 
Olivier Le Fevre\altaffilmark{29}, 
Benjamin Magnelli\altaffilmark{21}, 
Roberto Neri\altaffilmark{30}, 
Pascal Oesch\altaffilmark{31}, 
Kazuaki Ota\altaffilmark{32,5}, 
Hans--Walter Rix\altaffilmark{1}, 
Mark Sargent\altaffilmark{33}, 
Kartik Sheth\altaffilmark{34}, 
Arjen van der Wel\altaffilmark{1}, 
Paul van der Werf\altaffilmark{6}, 
Jeff Wagg\altaffilmark{35}
}
\altaffiltext{1}{Max-Planck Institut f\"{u}r Astronomie, K\"{o}nigstuhl 17, D-69117, Heidelberg, Germany. E-mail: {\sf decarli@mpia.de}}
\altaffiltext{2}{Astronomy Department, California Institute of Technology, MC105-24, Pasadena, California 91125, USA}
\altaffiltext{3}{National Radio Astronomy Observatory, Pete V.\,Domenici Array Science Center, P.O.\, Box O, Socorro, NM, 87801, USA}
\altaffiltext{4}{N\'{u}cleo de Astronom\'{\i}a, Facultad de Ingenier\'{\i}a, Universidad Diego Portales, Av. Ej\'{e}rcito 441, Santiago, Chile}
\altaffiltext{5}{Cavendish Laboratory, University of Cambridge, 19 J J Thomson Avenue, Cambridge CB3 0HE, UK}
\altaffiltext{6}{Leiden Observatory, Leiden University, PO Box 9513, NL2300 RA Leiden, The Netherland}
\altaffiltext{7}{Centre for Astrophysics and Supercomputing, Swinburne University of Technology, Hawthorn, Victoria 3122, Australia}
\altaffiltext{8}{Research School of Astronomy and Astrophysics, Australian National University, Canberra, ACT 2611, Australia}
\altaffiltext{9}{Laboratoire AIM, CEA/DSM-CNRS-Universite Paris Diderot, Irfu/Service d'Astrophysique, CEA Saclay, Orme des Merisiers, 91191 Gif-sur-Yvette cedex, France}
\altaffiltext{10}{European Southern Observatory, Karl-Schwarzschild-Strasse 2, 85748, Garching, Germany}
\altaffiltext{11}{Institute for Astronomy, University of Edinburgh, Royal Observatory, Blackford Hill, Edinburgh EH9 3HJ}
\altaffiltext{12}{Cornell University, 220 Space Sciences Building, Ithaca, NY 14853, USA}
\altaffiltext{13}{6 Centre for Extragalactic Astronomy, Department of Physics, Durham University, South Road, Durham, DH1 3LE, UK}
\altaffiltext{14}{Max-Planck-Institut f\"ur Radioastronomie, Auf dem H\"ugel 69, 53121 Bonn, Germany}
\altaffiltext{15}{Departamento de Ciencias F\'{\i}sicas, Universidad Andres Bello, Fernandez Concha 700, Las Condes, Santiago, Chile}
\altaffiltext{16}{Millennium Institute of Astrophysics (MAS), Nuncio Monse{\~{n}}or S{\'{o}}tero Sanz 100, Providencia, Santiago, Chile}
\altaffiltext{17}{Instituto de Astrof\'{\i}sica, Facultad de F\'{\i}sica, Pontificia Universidad Cat\'olica de Chile Av. Vicu\~na Mackenna 4860, 782-0436 Macul, Santiago, Chile}
\altaffiltext{18}{Millennium Institute of Astrophysics, Chile}
\altaffiltext{19}{Space Science Institute, 4750 Walnut Street, Suite 205, Boulder, CO 80301, USA}
\altaffiltext{20}{Department of Astronomy, University of Michigan, 1085 South University Ave., Ann Arbor, MI 48109, USA}
\altaffiltext{21}{Argelander Institute for Astronomy, University of Bonn, Auf dem H\"{u}gel 71, 53121 Bonn, Germany}
\altaffiltext{22}{Dalhousie University, Halifax, Nova Scotia, Canada}
\altaffiltext{23}{ASTRO-UAM, UAM, Unidad Asociada CSIC, Spain}
\altaffiltext{24}{Joint ALMA Observatory - ESO, Av. Alonso de C\'ordova, 3104, Santiago, Chile}
\altaffiltext{25}{National Radio Astronomy Observatory, 520 Edgemont Rd, Charlottesville, VA, 22903, USA}
\altaffiltext{26}{Steward Observatory, University of Arizona, 933 N. Cherry St., Tucson, AZ  85721, USA}
\altaffiltext{27}{Instituto de Astrof\'{\i}sica, Facultad de F\'{\i}sica, Pontificia Universidad Cat\'olica de Chile Av. Vicu\~na Mackenna 4860, 782-0436 Macul, Santiago, Chile}
\altaffiltext{28}{Instituto de F\'{\i}sica y Astronom\'{\i}a, Universidad de Valpara\'{\i}so, Avda. Gran Breta\~na 1111, Valparaiso, Chile}
\altaffiltext{29}{Aix Marseille Universite, CNRS, LAM (Laboratoire d'Astrophysique de Marseille), UMR 7326, F-13388 Marseille, France}
\altaffiltext{30}{IRAM, 300 rue de la piscine, F-38406 Saint-Martin d'H\`eres, France}
\altaffiltext{31}{Astronomy Department, Yale University, New Haven, CT 06511, USA}
\altaffiltext{32}{Kavli Institute for Cosmology, University of Cambridge, Madingley Road, Cambridge CB3 0HA, UK}
\altaffiltext{33}{Astronomy Centre, Department of Physics and Astronomy, University of Sussex, Brighton, BN1 9QH, UK}
\altaffiltext{34}{NASA Headquarters, Washington DC, 20546-0001, USA}
\altaffiltext{35}{SKA Organization, Lower Withington Macclesfield, Cheshire SK11 9DL, UK}

\begin{abstract}

In this paper we use ASPECS, the ALMA Spectroscopic Survey in the {\em Hubble} Ultra Deep Field (UDF) in band~3 and band~6, to place {\em blind constraints} on the CO luminosity function and the evolution of the cosmic molecular gas density as a function of redshift up to $z\sim 4.5$. This study is based on galaxies that have been solely selected through their CO emission and not through any other property. In all of the redshift bins the ASPECS measurements reach the predicted `knee' of the CO luminosity function (around $5\times10^{9}$\,\Kkmspc). We find clear evidence of an evolution in the CO luminosity function with respect to $z\sim 0$, with more CO luminous galaxies present at $z\sim 2$. The observed galaxies at $z\sim 2$ also appear more gas--rich than predicted by recent semi-analytical models. The comoving cosmic molecular gas density within galaxies as a function of redshift shows a factor 3--10 drop from $z \sim 2$ to $z \sim 0$ (with significant error bars), and possibly a decline at $z>3$. 
This trend is similar to the observed evolution of the cosmic star formation rate density. The latter therefore appears to be at least partly driven by the increased availability of molecular gas reservoirs at the peak of cosmic star formation ($z\sim2$).  
\end{abstract} \keywords{ galaxies: evolution ---
galaxies: ISM --- galaxies: star formation --- galaxies: statistics
--- submillimeter: galaxies --- instrumentation: interferometers}

\section{Introduction}

The cosmic star-formation history (SFH) describes the evolution of star formation in galaxies across cosmic time. It is well summarized by the so-called ``Lilly-Madau'' plot \citep{lilly95,madau96}, which shows the redshift evolution of the star-formation rate (SFR) density, i.e., the total SFR in galaxies in a comoving volume of the universe. The SFR density increases from an early epoch ($z>8$) up to a peak ($z\sim 2$) and then declines by a factor $\sim 20$ down to present day \citep[see][for a recent review]{madau14}. 

Three key quantities are likely to drive this evolution: the growth rate of dark matter halos, the gas content of galaxies (i.e., the availability of fuel for star formation), and the efficiency at which gas is transformed into stars. Around $z$=$2$, the mass of halos can grow by a factor of $>2$ in a Gyr; by $z\approx0$, the mass growth rate has dropped by an order of magnitude \citep[e.g.,][]{griffen16}. How does the halo growth rate affect the gas resupply of galaxies? Do galaxies at $z\sim 2$ harbor larger reservoirs of gas? Are they more effective at high redshift in forming stars from their gas reservoirs, possibly as a consequence of different properties of the interstellar medium, or do they typically have more disturbed gas kinematics due to gravitational interactions? 

To address some of these questions, we need a census of the dense gas stored in galaxies and available to form new stars as a function of cosmic time, i.e., the total mass of gas in galaxies per comoving volume [$\rho$(gas)]. The statistics of Ly$\alpha$ absorbers (associated with atomic hydrogen, H{\sc i}) along the line of sight toward bright background sources provide us with a measure of $\rho$(H{\sc i}). This appears to be consistent with being constant (within a $\sim$30\% fluctuation) from redshift $z=0.3$ to $z\sim 5$ \citep[see, e.g.,][]{crighton15}, possibly as a result of the balance between gas inflows and outflows in low-mass galaxies \citep{lagos14} and of the on-going gas resupply from the intergalactic medium \citep{lagos11}. However, beyond the local universe, little information currently exists on the amount of molecular gas  that is stored in galaxies, $\rho$(H$_2$), which is the immediate fuel for star formation (e.g., see review by \citealt{carilli13}).

Attempts have been made to infer the mass of molecular gas in distant targeted galaxies indirectly from the measurement of their dust emission, via dust--to--gas scaling relations \citep{magdis11,magdis12,scoville14,scoville15,groves15}. But a more direct route is to derive it from the observations of rotational transitions of $^{12}$CO (hereafter, CO), the second most abundant molecule in the universe (after H$_2$). As the second approach is most demanding in terms of telescope time, it has been traditionally applied only with extreme, infrared (IR) luminous sources (e.g., \citealt{bothwell13}; these however account for only 10-20\% of the total SFR budget in the universe; see, \citealt{rodighiero11,magnelli13,gruppioni13,casey14}), or on samples of galaxies pre-selected based on their stellar mass and/or SFR \citep[e.g.,][]{daddi10a,daddi10b,daddi15,tacconi10,tacconi13,genzel10,genzel15,bolatto15}. These observations have been instrumental in shaping our understanding of the molecular gas properties in high-$z$ galaxies. 
Through the observation of multiple CO transitions for single galaxies, the CO excitation has been constrained in a variety of systems \citep{weiss07,riechers11,bothwell13,spilker14,daddi15}. Most remarkably, various studies showed that $M_*$- and SFR-selected galaxies at $z>0$ tend to host much larger molecular gas reservoirs than typically observed in local galaxies for a given stellar mass ($M_*$) suggesting that an evolution in the gas fraction $f_{\rm gas}=M_{\rm H2}/(M_*+M_{\rm H2})$ occurs through cosmic time \citep{daddi10a,riechers10,tacconi10,tacconi13,genzel10,genzel15,geach11,magdis12,magnelli12}.


For molecular gas observations to constrain $\rho$(H$_2$) as a function of cosmic time, we need to sample the CO luminosity function in various redshift bins. CO is the second most abundant molecule in the universe (after H$_2$) and therefore is an excellent tracer of the molecular phase of the gas. The CO(1-0) ground transition has an excitation temperature of only $T_{\rm ex}=5.5$\,K, i.e., the molecule is excited in virtually any galactic environment. Other low-J CO lines may be of practical interest, as these levels remain significantly excited in star-forming galaxies; and thus, the associated lines [CO(2-1), CO(3-2), CO(4-3)] are typically brighter and easier to detect than the ground state transition CO(1-0). There have been various predictions of the CO luminosity functions both for the J=1$\rightarrow$0 transition and for intermediate and high-J lines, using either theoretical models \citep[e.g.,][]{obreschkow09a,obreschkow09b,lagos11,lagos12,lagos14,popping14a,popping14b,popping16} or empirical relations \citep[e.g.,][]{sargent12,sargent14,dacunha13,vallini16}.

Theoretical models typically rely on semi-analytical estimates of the budget of gas in galaxies (e.g.,  converting H{\sc i} into H$_2$ assuming a pressure-based argument, as in \citealt{blitz06}; via metallicity-based arguments, as in \citealt{gnedin10,gnedin11}; or based on the intensity of the radiation field and the gas properties, as in \citealt{krumholz08,krumholz09}), and inferring the CO luminosity and excitation via radiative transfer models. These models broadly agree on the dependence of $\rho$(H$_2$) on $z$, at least up to $z\sim 2$, but widely differ in the predicted CO luminosity functions, in particular for intermediate and high J transitions, where details on the treatment of the CO excitation become critical. For example, the models by \citet{lagos12} predict that the knee of the CO(4-3) luminosity function lies at $L'\approx 5\times10^8$\,\Kkmspc{} at $z\sim 3.8$, while the models by \citet{popping16} place the knee at a luminosity about 10 times brighter. Such a spread in the predictions highlight the lack of observational constraints to guide the theoretical assumptions.

This study aims at providing observational constraints on the CO luminosity functions and cosmic density of molecular gas via the `molecular deep field' approach. We perform a scan over a large range of frequency ($\Delta \nu/\nu\approx25-30$\,\%) in a region of the sky, and ``blindly'' search for molecular gas tracers at any position and redshift. By focusing on a blank field, we avoid the biases due to pre-selection of sources. This method naturally provides us with a well-defined cosmic volume where to search for CO emitters, thus leading to direct constraints on the CO luminosity functions. Our first pilot experiment with the IRAM Plateau de Bure Interferometer \citep[PdBI; see][]{decarli14} led to the first, weak constraints on the CO luminosity functions at $z>0$ \citep{walter14}. The modest sensitivity (compared with the expected knee of the CO luminosity functions) resulted in large Poissonian uncertainties. These can be reduced now, thanks to the Atacama Large Millimeter/Sub-millimeter Array (ALMA). 

We obtained ALMA Cycle 2 observations to perform two spatially coincident molecular deep fields, at 3mm and 1mm respectively, in a region of the Hubble Ultra Deep Field \citep[UDF,][]{beckwith06}. The data set of our ALMA Spectroscopic Survey (ASPECS) is described in detail in Paper I of this series \citep{walter16}. Compared with the aforementioned PdBI effort, we now reach a factor of 3--4 better sensitivity, which allows us to sample the expected knee of the CO luminosity functions over a large range of transitions. Furthermore, the combination of band 3 and 6 offers us direct constraints on the CO excitation of the observed sources, thus allowing us to infer the corresponding CO(1-0) emission, and therefore $\rho$(H$_2$). The collapsed cube of the 1mm observations also yields one of the deepest dust continuum observations ever obtained \citep[Paper II of this series,][]{aravena16a}, which  we can use to compare the $\rho$(H$_2$) estimates based on CO and the $\rho$(gas) estimates based on the dust emission. 

This paper is organized as follows: In Sec.~\ref{sec_observations} we summarize the observations and the properties of the data set. In Sec.~\ref{sec_analysis} we describe how we derive our constraints on the CO luminosity functions and on $\rho$(H$_2$) and $\rho$(gas). In Sec.~\ref{sec_discussion} we discuss our results. Throughout the paper we assume a standard $\Lambda$CDM cosmology with $H_0=70$ km s$^{-1}$ Mpc$^{-1}$, $\Omega_{\rm m}=0.3$ and $\Omega_{\Lambda}=0.7$ \citep[broadly consistent with the measurements by the][]{planck15}.


\section{Observations}\label{sec_observations}


The data set used in this study consists of two frequency scans at 3mm (band 3) and 1mm (band 6) obtained with ALMA in the UDF centered at RA = 03:32:37.900, Dec = --27:46:25.00 (J2000.0). Details on the observations and data reduction are presented in Paper I, but the relevant information is briefly summarized here. The 3mm scan covers the 84-115 GHz range with a single spatial pointing. The primary beam of the 12m ALMA antennas is $\sim75''$ at 84\,GHz and $\sim 54''$ at 115\,GHz. The typical RMS noise is 0.15\,mJy\,beam$^{-1}$ per 20\,MHz channel. The 1mm scan encompasses the frequency window 212-272 GHz. In order to sample a similar area as in the 3mm scan, given the smaller primary beam ($\sim 26''$), we performed a 7 point mosaic. The typical depth of the data is $\sim 0.5$\,mJy\,beam$^{-1}$ per 30 MHz channel. The synthesized beams are $\sim 3.5''\times 2.0''$ at 3mm and $\sim 1.5''\times 1.0''$ at 1mm. 

\begin{figure}
\includegraphics[width=0.99\columnwidth]{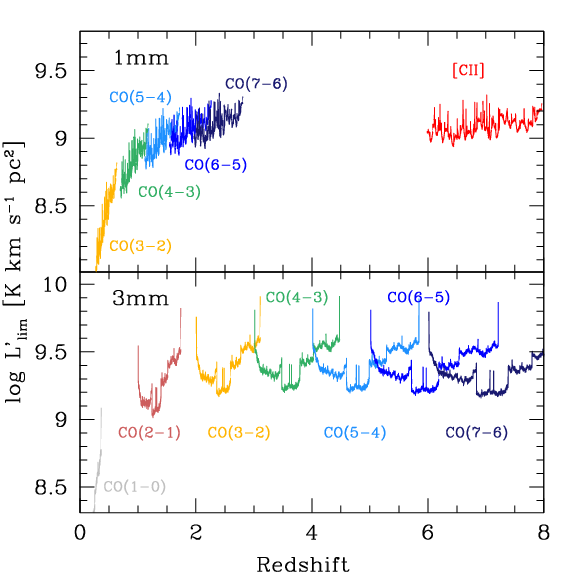}\\
\caption{Redshift coverage and luminosity limit reached in our 1mm and 3mm scans, for various CO transitions and for the \Cii{} line. The (5-$\sigma$) limits plotted here are computed assuming point-source emission, and are based on the observed noise per channel, scaled for a line width of 200 \kms{}. The combination of band 3\&6 offers a virtually-complete CO redshift coverage. The luminosity limit (expressed as velocity-integrated temperature) is roughly constant at $z\gsim 1.5$. The depth of our observations is sufficient to sample the typical knee of the expected CO luminosity functions ($L'\sim5\times10^9$\,\Kkmspc{}). }
\label{fig_lum_lim}
\end{figure}

Fig.~\ref{fig_lum_lim} shows the redshift ranges and associated luminosity limits reached for various transitions in the two bands. The combination of band 3 and band 6 provides virtually-complete CO redshift coverage. The luminosity limits are computed assuming 5-$\sigma$ significance, a line width of 200\,\kms{}, and unresolved emission at the angular resolution of our data. At $z\gsim 1.5$, the luminosity limit (expressed as a velocity-integrated temperature over the beam, which is constant for all CO transitions in the case of thermalized emission) is roughly constant as a function of redshift for different CO transitions as well as for \Cii{}: $\sim 2\times 10^9$\,\Kkmspc{}. 

\section{Analysis}\label{sec_analysis}

Given the blank field approach of ASPECS, with no pre-selection on the targeted sources, we have a well-defined, volume-limited sample of galaxies at various redshifts where we can search for CO emission. We first concentrated on the ``blind'' CO detections presented in Paper I (Tab.~2), and then include the information from galaxies with a known redshift. This provides us with direct constraints on the CO luminosity function in various redshift bins. We then use these constraints to infer the CO(1-0) luminosity functions in various redshift bins, and therefore the H$_2$ mass ($M_{\rm H2}$) budget in galaxies throughout cosmic time.

\subsection{CO detections}

\subsubsection{Blind detections}\label{sec_blind}

\begin{table*}
\caption{{\rm Catalogue of the line candidates discovered with the blind line search. 
(1) Line ID. (2-3) Right ascension and declination (J2000).  
(4) Fidelity level at the S/N of the line candidate.
(5) Completeness at the luminosity of the line candidate.
(6) Is there an optical/near-IR counterpart? 
(7) Notes on line identification: 
   {\em i}- Multiple lines detected in the ASPECS cubes; 
   {\em ii}- Lack of other lines in the ASPECS cubes; 
   {\em iii}- Absence of optical/near-IR counterpart suggests high $z$;
   {\em iv}- Supported by (a) spectroscopic, (b) grism, or (c) photometric redshift.
(8) Possible line identification. A cardinal number indicates the upper J level of a CO transition. 
(9) CO redshift corresponding to the adopted line identification. 
(10) Line luminosity, assuming the line identification in col.(8). The uncertainties are propagated from the uncertainties in the line flux measurement. 
(11) Molecular gas mass $M_{\rm H2}$ as derived from the observed CO luminosity (see eq.~\ref{eq_H2}), only for J$<$5 CO lines. 
}} \label{tab_lines}
\begin{center}
\begin{tabular}{ccccccccccc}
\hline
ASPECS ID & RA	  & Dec	        & Fid.     & $C$  & C.part? & Notes & Line & $z_{\rm CO}$ & $L'$          & $M_{\rm H2}$  \\
    &             &             &         &      &         &       &ident.&              &[$10^8$\Kkmspc]& [$10^8$\Msun]  \\
 (1)& (2)         & (3)         & (4)     & (5)  & (6)     & (7)   & (8)  & (9)          & (10)          & (11)              \\
\hline
\multicolumn{11}{c}{3mm}\\
3mm.1     & 03:32:38.52 & --27:46:34.5 & 1.00 & 1.00 & Y & {\em i, iv}(b)  &	    3	& 2.5442 & $240.4\pm1.0$  & $2061\pm 9$   \\   
3mm.2     & 03:32:39.81 & --27:46:11.6 & 1.00 & 1.00 & Y & {\em i, iv}(a)  &	    2	& 1.5490 & $136.7\pm2.1$  & $ 648\pm10$   \\   
3mm.3     & 03:32:35.55 & --27:46:25.7 & 1.00 & 0.85 & Y & {\em iv}(a)	   &	    2	& 1.3823 &  $33.7\pm0.7$  & $ 160\pm 3$   \\   
3mm.4     & 03:32:40.64 & --27:46:02.5 & 1.00 & 0.85 & N & {\em ii}        &	    3	& 2.5733 &  $45.8\pm1.0$  & $ 393\pm 9$   \\   
          &             &              &      &      &   &                 &	    4	& 4.0413 &  $92.2\pm2.8$  & $1071\pm33$   \\   
          &             &              &      &      &   &                 &	    5	& 5.3012 &  $89.5\pm2.7$  & 	---       \\   
3mm.5     & 03:32:35.48 & --27:46:26.5 & 0.87 & 0.85 & Y & {\em iv}(a)	   &	    2	& 1.0876 &  $28.3\pm0.9$  & $ 134\pm 4$   \\   
3mm.6     & 03:32:35.64 & --27:45:57.6 & 0.86 & 0.85 & N & {\em ii, iii}   &	    3	& 2.4836 &  $72.8\pm1.0$  & $ 624\pm 9$   \\   
          &             &              &      &      &   &                 &	    4	& 3.6445 &  $77.3\pm1.0$  & $ 898\pm12$   \\   
          &             &              &      &      &   &                 &	    5	& 4.8053 &  $76.2\pm1.0$  & 	---       \\   
3mm.7     & 03:32:39.26 & --27:45:58.8 & 0.86 & 0.85 & N & {\em ii, iii}   &	    3	& 2.4340 &  $25.9\pm1.0$  & $ 222\pm 9$   \\   
          &             &              &      &      &   &                 &	    4	& 3.5784 &  $27.6\pm1.0$  & $ 321\pm12$   \\   
          &             &              &      &      &   &                 &	    5	& 4.7227 &  $27.3\pm1.0$  & 	---       \\   
3mm.8     & 03:32:40.68 & --27:46:12.1 & 0.76 & 0.85 & N & {\em ii, iii}   &	    3 	& 2.4193 &  $58.6\pm0.9$  & $ 502\pm 8$   \\   
          &             &              &      &      &   &                 &	    4	& 3.5589 &  $62.6\pm1.0$  & $ 727\pm12$   \\   
          &             &              &      &      &   &                 &	    5	& 4.6983 &  $62.0\pm1.0$  & 	---       \\   
3mm.9     & 03:32:36.01 & --27:46:47.9 & 0.74 & 0.85 & N & {\em ii, iii}   &	    3	& 2.5256 &  $30.5\pm1.0$  & $ 261\pm 9$   \\   
          &             &              &      &      &   &                 &	    4	& 3.7006 &  $32.3\pm1.0$  & $ 375\pm12$   \\   
          &             &              &      &      &   &                 &	    5	& 4.8754 &  $31.8\pm1.0$  & 	---       \\   
3mm.10    & 03:32:35.66 & --27:45:56.8 & 0.61 & 0.85 & Y & {\em ii, iv}(b) &	    3	& 2.3708 &  $70.4\pm0.9$  & $ 603\pm 8$   \\   
\hline														    	    	      
\multicolumn{11}{c}{1mm}\\
1mm.1$^*$ & 03:32:38.54 & --27:46:34.5 & 1.00 & 1.00 & Y & {\em i, iv}(b)  &	  7	& 2.5439 & $48.02\pm0.37$ & 	---       \\   
1mm.2$^*$ & 03:32:38.54 & --27:46:34.5 & 1.00 & 1.00 & Y & {\em i, iv}(a)  &	  8	& 2.5450 & $51.42\pm0.23$ & 	---       \\   
1mm.3     & 03:32:38.54 & --27:46:31.3 & 0.93 & 0.85 & Y & {\em iv}(b)	   &	  3	& 0.5356 &  $3.66\pm0.08$ &  $ 31\pm1$    \\   
1mm.4     & 03:32:37.36 & --27:46:10.0 & 0.85 & 0.65 & N & {\em i}	   &   \Cii	& 6.3570 & $12.49\pm0.23$ & 	---       \\   
1mm.5     & 03:32:38.59 & --27:46:55.0 & 0.79 & 0.75 & N & {\em ii}	   &     4	& 0.7377 & $12.95\pm0.09$ &  $150\pm1$    \\   
          &             &              &      &      &   &                 &   \Cii	& 6.1632 & $31.84\pm0.22$ & 	---       \\   
1mm.6     & 03:32:36.58 & --27:46:50.1 & 0.78 & 0.75 & Y & {\em iv}(c)	   &     4	& 1.0716 & $21.45\pm0.15$ &  $249\pm2$    \\   
          &             &              &      &      &   &                 &     5	& 1.5894 & $29.12\pm0.21$ & 	---       \\   
          &             &              &      &      &   &                 &     6	& 2.1070 & $33.68\pm0.24$ & 	---       \\   
1mm.7     & 03:32:37.91 & --27:46:57.0 & 0.77 & 1.00 & N & {\em ii, iii}   &     4	& 0.7936 & $37.53\pm0.10$ &  $436\pm1$    \\   
          &             &              &      &      &   &                 &   \Cii	& 6.3939 & $84.01\pm0.23$ & 	---       \\   
1mm.8     & 03:32:37.68 & --27:46:52.6 & 0.71 & 0.72 & N & {\em ii, iii}   &   \Cii	& 7.5524 & $23.22\pm0.24$ & 	---       \\   
1mm.9     & 03:32:36.14 & --27:46:37.0 & 0.63 & 0.75 & N & {\em ii, iii}   &     4	& 0.8509 &  $8.21\pm0.12$ &  $ 95\pm1$    \\   
          &             &              &      &      &   &                 &   \Cii	& 6.6301 & $16.84\pm0.25$ & 	---       \\   
1mm.10    & 03:32:37.08 & --27:46:19.9 & 0.62 & 0.75 & N & {\em ii, iii}   &     4	& 0.9442 & $14.74\pm0.18$ &  $171\pm2$    \\   
          &             &              &      &      &   &                 &     6	& 1.9160 & $25.05\pm0.30$ & 	---       \\   
          &             &              &      &      &   &                 &   \Cii	& 7.0147 & $26.59\pm0.32$ & 	---       \\   
1mm.11    & 03:32:37.71 & --27:46:41.0 & 0.61 & 0.85 & N & {\em ii, iii}   &     3	& 0.5502 &  $4.84\pm0.09$ &  $ 41\pm1$    \\   
          &             &              &      &      &   &                 &   \Cii	& 7.5201 & $16.25\pm0.30$ & 	---       \\   
\hline
\end{tabular}
\begin{tablenotes}
      \small
      \item $^*$ Not used for deriving the H$_2$ mass for this source, as a lower-J line is available.
\end{tablenotes}
\end{center}
\end{table*}								 

In Paper I, we describe our ``blind search'' of CO emission purely based on the ALMA data (i.e., with no support from ancillary data at other wavelengths)\footnote{The code for the blind search of line candidates is publicly available at \textsf{http://www.mpia.de/homes/decarli/ASPECS/findclumps.cl}.}. In brief, we perform a floating average of consecutive frequency channels in bins of $\sim$50--300\,\kms{} in the imaged cubes. For each averaged image, we compute the map rms and select peaks based on their S/N. A search for negative (= noise) peaks allows us to quantify the fidelity of our line candidates based on their S/N, and the injection of mock lines allows us to assess the level of completeness of our search as a function of various line parameters, including the line luminosity. The final catalogue consists of 10 line candidates from the 3mm cube, and 11 from the 1mm cube. We use a Gaussian fit of the candidate spectra to estimate the line flux, width, and frequency (see Tab.~2 of Paper I), and we investigate the available optical/near-IR images to search for possible counterparts.

The line identification (and therefore, the redshift association) requires a number of stpng, similar to our earlier study of the HDF--N \citep{decarli14}, which are as follows:
\begin{itemize}
\item[{\em i-}] We inspect the cubes at the position of each line candidate, and search for multiple lines. If multiple lines are found, the redshift should be uniquely defined. Since $\nu_{\rm CO[J-(J-1)]}\approx {\rm J}\,\nu_{\rm CO(1-0)}$, some ambiguity may still be in place [e.g., two lines with a frequency ratio of 2 could be CO(2-1) and CO(4-3), or CO(3-2) and CO(6-5)]. In these cases, the following stpng allow us to break the degeneracy.
\item[{\em ii-}] The absence of multiple lines can then be used to exclude some redshift identification. E.g., lines with similar J should show similar fluxes, under reasonable excitation conditions. If we identify a bright line as, e.g., CO(5-4), we expect to see a similarly-luminous CO(4-3) line (if this falls within the coverage of our data set). If that is not the case, we can exclude this line identification.
\item[{\em iii-}] The exquisite depth of the available multi-wavelength data allows us to detect the starlight emission of galaxies with stellar mass $M_{*}\sim10^8$\,\Msun{} at almost all $z<2$. In the absence of an optical / near-IR counterpart, we thus exclude redshift identification that would locate the source at $z<2$.
\item[{\em iv-}] In the presence of an optical/near-IR counterpart, the line identification is guided by the availability of optical redshift estimates. Optical spectroscopy \citep[e.g., see the compilations by][]{lefevre05,coe06,skelton14,morris15} is considered secure (typical uncertainties are in the order of a few hundred \kms{}). When not available, we rely on {\em HST} grism data \citep{morris15,momcheva16}, or photometric redshifts \citep{coe06,skelton14}.
\end{itemize}
Ten out of 21 blindly-selected lines are uniquely identified in this way. A bootstrap analysis is then adopted to account for the remaining uncertainties in the line identification: To each source, we assign a redshift probability distribution which is proportional to the comoving volume in the redshift bins sampled with all the possible line identifications. We then run 1000 extractions of the redshift values picked from their probability distributions and compute the relevant quantities (line luminosities, inferred molecular masses, contribution to the cosmic density of molecular gas) in each case. The results are then averaged among all the realizations. The line identifications and associated redshifts are listed in Tab.~\ref{tab_lines}.

\begin{figure*}
\includegraphics[width=0.99\columnwidth]{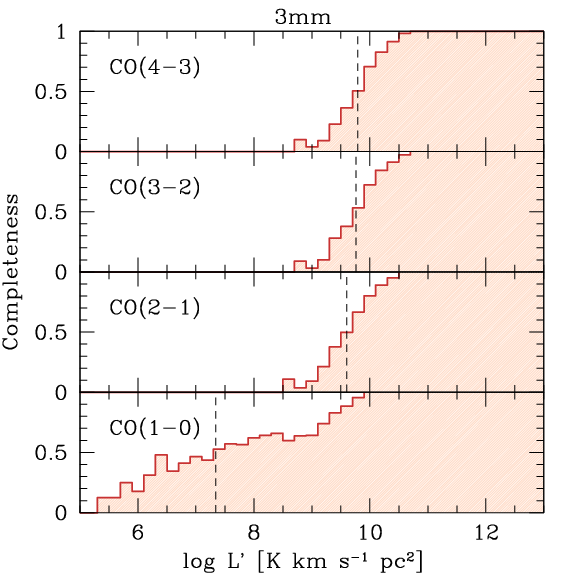}
\includegraphics[width=0.99\columnwidth]{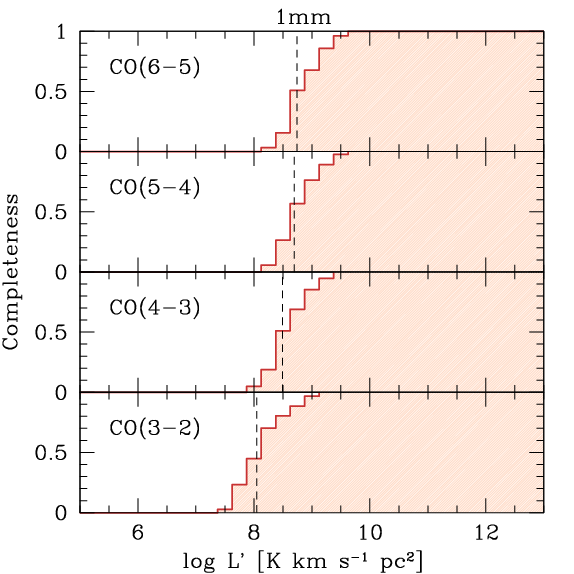}\\
\caption{Luminosity limit reached in our 3mm and 1mm scans, for various CO transitions. The completeness is computed as the number of mock lines retrieved by our blind search analysis divided by the number of ingested mock lines, and here it is plotted as a function of the line luminosity. The 50\% limits, marked as dashed vertical lines, are typically met at $L'=(3-6) \times 10^9$\,\Kkmspc{} at 3mm for any J$>$1, and at $L'=(4-8) \times 10^8$\,\Kkmspc{} at 1mm for any J$>$3. The J=1 and 3 cases in the 3mm and 1mm cubes show a broader distribution towards lower luminosity limits due to the wide spread of luminosity distance for these transitions within the frequency ranges of our observations.}
\label{fig_completeness}
\end{figure*}
To compute the contribution of each line candidate to the CO luminosity functions and to the cosmic budget of molecular gas mass in galaxies, we need to account for the fidelity (i.e., the reliability of a line candidate against false-positive detections) and completeness (i.e., the fraction of line candidates that we retrieve as a function of various line parameters) of our search. For the fidelity, we infer the incidence of false-positive detections from the statistics of negative peaks in the cubes as a function of the line S/N, as described in Sec.~3.1.1 of Paper I. Fig.~\ref{fig_completeness} shows the completeness of our line search as a function of the line luminosity. This is obtained by creating a sample of 2500 mock lines (as point sources), with a uniform distribution of frequency, peak flux density, width, and position within the primary beam. Under the assumption of observing a given transition [e.g., CO(3-2)], we convert the input frequency into redshift, and the integrated line flux ($F_{\rm line}$) from the peak flux density and width. We then compute line luminosities for all the mock input lines as:
\begin{equation}\label{eq_L1def}
\frac{L'}{\rm K\,km\,s^{-1}\,pc^2}=\frac{3.25\times 10^7}{(1+z)^3} \, \frac{F_{\rm line}}{\rm Jy\,km\,s^{-1}} \left(\frac{\nu_{\rm obs}}{\rm GHz}\right)^{-2} \left(\frac{D_{\rm L}}{\rm Mpc}\right)^2
\end{equation}
where $\nu_{\rm obs}$ is the observed frequency of the line, and $D_{\rm L}$ is the luminosity distance \citep[see, e.g.,][]{solomon97}. Finally, we run our blind line search algorithm and display the fraction of retrieved-to-input lines as a function of the input line luminosity. Our analysis is 50\% complete down to line luminosities of $(4-6) \times 10^9$\,\Kkmspc{} at 3mm for any J$>$1, and $(1-6) \times 10^8$\,\Kkmspc{} at 1mm for any J$>$3, in the area corresponding to the primary beam of the 3mm observations. The completeness distributions as a function of line luminosity in the J=1 case (at 3mm) and the J=3 case (at 1mm) show long tails towards lower luminosities due to the large variations of $D_{\rm L}$ within our scans for these lines (see also Fig.~\ref{fig_lum_lim}). 
The levels of fidelity and completeness at the S/N and luminosity of the line candidates in our analysis is reported in Tab.~\ref{tab_lines}. At low S/N, flux boosting might bias our results high, through effectively overestimating the impact of a few intrinsically bright sources against many fainter ones scattered above our detection threshold by the noise. However, the relatively high S/N ($>$5) of our line detections, and the statistiscal corrections for missed lines that are scattered below our detection threshold, and for spurious detections, make the impact of flux boosting negligible in our analysis.

\subsubsection{CO line stack}

\begin{figure}
\includegraphics[width=0.99\columnwidth]{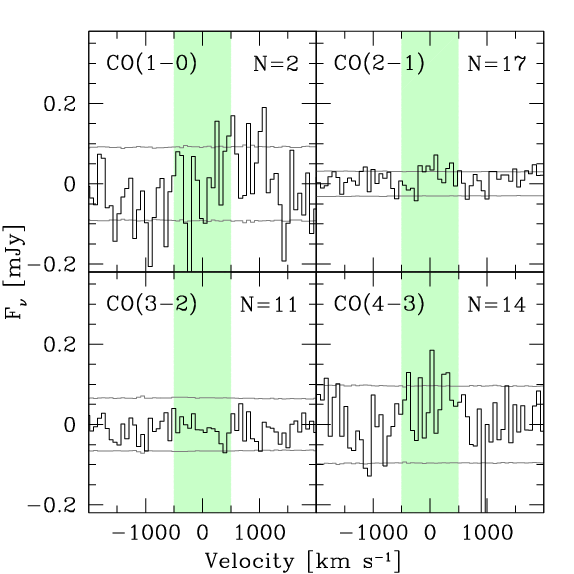}\\
\caption{Stacked mm spectrum of the sources in our field with optical/near-IR redshifts. The adopted spectral bin is 70\,\kms{} wide. The 1-$\sigma$ uncertainties are shown as grey lines. We highlight the $\pm$500\,\kms{} range where the stacked flux is integrated. We also list the number of sources entering each stack. No clear detection is reported in any of the stacked transitions.}
\label{fig_zspec_stack}
\end{figure}

We can improve the sensitivity of our CO search beyond our `blind' CO detections by focusing on those galaxies where an accurate redshift is available via optical/near-IR spectroscopy. Slit spectroscopy typically leads to uncertainties of a few 100 \kms{}, while grism spectra from the 3D-{\em HST} \citep{momcheva16} have typical uncertainties of $\sim 1000$ \kms{} due to the coarser resolution and poorer S/N. By combining the available spectroscopy, we construct a list of 42 galaxies for which a slit or grism redshift information is available \citep{lefevre05,coe06,skelton14,morris15,momcheva16} within $37.5''$ from our pointing center (this corresponds to the area of the primary beam at the low-frequency end of the band 3 scan). Out of these, 36 galaxies have a redshift for which one or more J$<$5 CO transitions have been covered in our frequency scans. We extract the 3mm and 1mm spectra of all these sources, and we stack them with a weighted-average. As weights, we used the inverse of the variance of the spectral noise. This is the pixel rms of each channel map, corrected a posteriori for the primary beam attenuation at the source position. As Fig.~\ref{fig_zspec_stack} shows, no obvious line is detected above a S/N=3. If we integrate the signal in a 1000\,\kms{} wide bin centered on the rest-frame frequency of the lines, we retrieve a $\sim 2$-$\sigma$ detection of the CO(2-1) and CO(4-3) lines (corresponding to average line fluxes of $\sim 0.006$\,Jy\,\kms{} and $\sim 0.010$\,Jy\,\kms{} respectively). However, given that their low significance, and that they are drawn from a relatively sparse sample, we opt not to include them in the remainder of the analysis, until we are able to significantly expand the list of sources with secure optical/near-IR redshifts. This will be possible thanks to the advent of integral field spectroscopy units with large field of view, like MUSE, which will provide spectra (and therefore redshifts) for hundreds of galaxies in our pointing. 

\subsection{CO luminosity functions}

The CO luminosity functions are constructed as follows:
\begin{equation}\label{eq_LF}
\Phi (\log L_i)=\frac{1}{V}\,\sum_{j=1}^{N_i}\frac{{\rm Fid}_j}{C_j}
\end{equation}
Here, $N_i$ is the number of galaxies with a CO luminosity falling into the luminosity bin $i$, defined as the luminosity range between $\log L_i-0.5$ and $\log L_i+0.5$, while $V$ is the volume of the universe sampled in a given transition. Each entry $j$ is down-weighted according to the fidelity (Fid$_j$) and up-scaled according to the completeness ($C_j$) of the $j$-th line. As described in Paper I, the fidelity at a given S/N is defined as $(N_{\rm pos}-N_{\rm neg})/N_{\rm pos}$, where $N_{\rm pos/neg}$ is the number of positive and negative lines with said S/N. This definition of the fidelity allows us to statistically subtract the false positive line candidates from our blind selection. The uncertainties on $\Phi(\log L_i)$ are set by the Poissonian errors on $N_i$, according to \citet{gehrels86}\footnote{According to \citet{cameron11}, the binomial confidence intervals in \citet{gehrels86} might be overestimated in the low-statistics regime compared to a fully Bayesian treatment of the distributions. A similar effect is possibly in place for Poissonian distributions, although a formal derivation is beyond the scope of this work. Here we conservatively opt to follow the classical \citet{gehrels86} method.}. We consider the confidence level corresponding to 1-$\sigma$.  We include the uncertainties associated with the line identification and the errors from the flux measurements in the bootstrap analysis described in Sec.~\ref{sec_blind}. Given that all our blind sources have S/N$>$5 by construction, and the number of entries is typically of a few sources per bin, Poissonian uncertainties always dominate. The results of the bootstrap are averaged in order to produce the final luminosity functions.

\begin{figure}
\includegraphics[width=0.99\columnwidth]{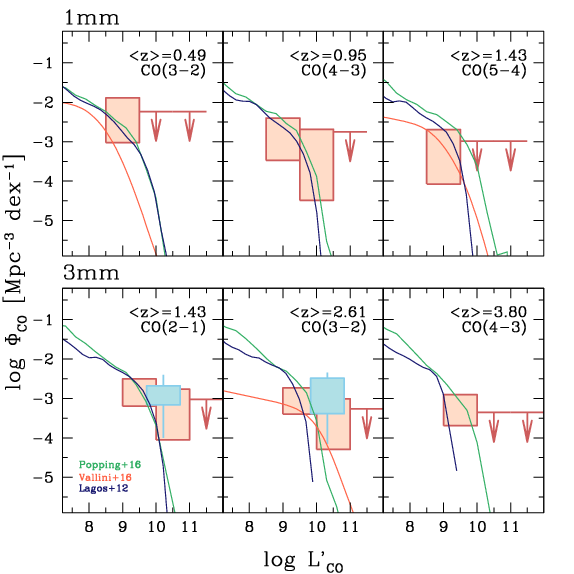}\\
\caption{CO luminosity functions in various redshift bins. The constraints from our ALMA UDF project are marked as red squares, with the vertical size of the box showing the Poissonian uncertainties. The results of the HDF study by \citet{walter14} are shown as cyan boxes, with error bars marking the Poissonian uncertainties. Semianalytical models by \citet{lagos12} and \citet{popping16} as well as the empirical predictions by \citet{vallini16} are shown for comparison. Our ALMA observations reach the depth required to sample the expected knee of the luminosity functions in most cases \citep[the only exception being the $\langle z \rangle=3.80$ bin when compared with the predictions by][]{lagos12}. Our observations reveal an excess of CO-luminous sources at the bright end of the luminosity function, especially in the 3mm survey, with respect to the predictions. Such an excess is not observed in the 1mm, suggesting that the CO excitation is typically modest compared to the models shown here.}
\label{fig_co_lf}
\end{figure}

The CO luminosity functions obtained in this way are shown in Fig.~\ref{fig_co_lf}. For comparison, we include the predictions based on semi-analytical models by \citet{lagos12} and \citet{popping16} and on empirical IR luminosity function of {\em Herschel} sources by \citet{vallini16}, as well as the constraints obtained by the earlier study of the HDF--N \citep{walter14}.
\begin{figure*}
\includegraphics[width=0.99\textwidth]{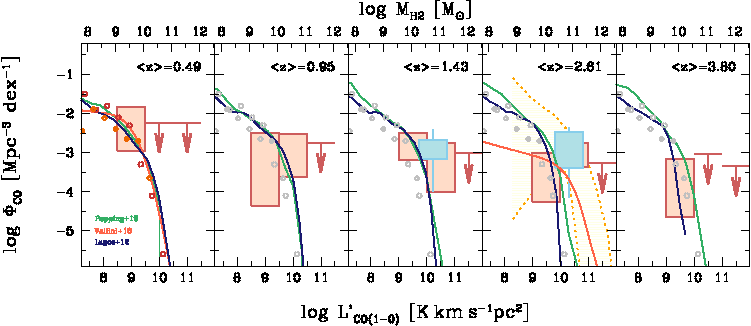}\\
\caption{CO(1-0) luminosity functions in various redshift bins. The constraints from ASPECS are marked as red squares, with the vertical size of each box showing the uncertainties. The results from the 3mm scan with PdBI by \citet{walter14} are shown as cyan boxes, with error bars marking the Poissonian uncertainties. The observed CO(1-0) luminosity functions of local galaxies by \citet{keres03} and \citet{boselli14} are shown as red circles and orange diamonds in the first panel, respectively, and as grey points for comparison in all the other panels. The intensity mapping constraints from \citet{keating16} are shown as a shaded yellow area. Semi-analytical models by \citet{lagos12} and \citet{popping16} as well as the empirical predictions by \citet{vallini16} are shown for comparison. The mass function scale shown in the top assumes a fixed $\alpha_{\rm CO}=3.6$\,\Msun{}(\Kkmspc)$^{-1}$. Our results agree with the predictions at $z<1$, and suggest that an excess of bright sources with respect to both the empirical predictions by \citet{vallini16} and the models by \citet{lagos12} appears at $z>1$.}
\label{fig_co10_lf}
\end{figure*}
Our observations reach the knee of the luminosity functions in almost all redshift bins. The only exception is the CO(4-3) transition in the $\langle z \rangle=3.80$ bin, for which the models by \citet{lagos12} place the knee approximately one order of magnitude below that predicted by \citet{popping16}, thus highlighting the large uncertainties in the state-of-the-art predictions of gas content and CO excitation, especially at high redshift. In particular, these two approaches differ in the treatment of the radiative transfer and CO excitation in a number of ways: 1) \citet{lagos12} adopt a single gas density value for each galaxy, whereas \citet{popping16} construct a density distribution for each galaxy, and assume a log-normal density distribution for the gas within clouds; 2) \citet{lagos12} include heating from both UV and X-rays (although the latter might be less critical for the purposes of this paper), while \citet{popping16} only consider the UV contribution to the heating; 3) the CO chemistry in \citet{lagos12} is set following the \textsf{UCL\_PDR} photo-dissociation region code \citep{bell06,bell07}, and in \citet{popping16} it is based on a fit to results from the \citet{wolfire10} photo-dissociation region code; 4) the CO excitation in \citet{lagos12} is also based on the \textsf{UCL\_PDR} code, while \citet{popping16} adopt a customized escape probability code for the level population; 5) the typical $\alpha_{\rm CO}$ in the \citet{lagos12} models is higher than in \citet{popping16}, although the exact value of $\alpha_{\rm CO}$ in both models changes from galaxy to galaxy [i.e., the CO(1-0) luminosity functions do not translate into H$_2$ mass functions with a simple scaling].

Our observations shown in Fig.~\ref{fig_co_lf} indicate that an excess of CO-bright sources with respect to semi-analytical models might be in place. This is apparent in the 3mm data. However, the same excess is not observed in the 1mm band. In particular, in the $\langle z \rangle = 1.43$ bin, the lack of bright CO(5-4) lines [compared to the brighter CO(2-1) emission reported here] suggests that the CO excitation is typically modest.

Such apparent low CO excitation is supported by the detailed analysis of a few CO-bright sources presented in a companion paper \citep[Paper IV of this series,][]{decarli16}. These findings guide our choice of a low-excitation template to convert the observed J$>$1 luminosities into CO(1-0). In the next stpng of our analysis, we refer to the template of CO excitation of main sequence galaxies by \citet{daddi15}: If $r_{\rm J1}$ is the temperature ratio between the CO(J-[J-1]) and the CO(1-0) transitions, we adopt $r_{\rm J1}$=$0.76\pm0.09$, $0.42\pm0.07$, $0.23\pm0.04$ for J=2,3,5. In the case of CO(4-3) (which is not part of the template), we interpolate the models shown in the left-hand panel of Fig.~10 in \citet{daddi15}, yielding $r_{41}$=$0.31\pm0.06$, where we conservatively assume a 20\% uncertainty. Each line luminosity is then converted into CO(1-0) as:
\begin{equation}\label{eq_co10}
\log L_{\rm CO(1-0)}'=\log L_{\rm CO(J-[J-1])}'-\log r_{\rm J1}
\end{equation}
The uncertainties in the excitation correction are included in the bootstrap analysis described in Sec.~\ref{sec_blind}. Based on these measurements, we derive CO(1-0) luminosity functions following eq.~\ref{eq_LF}. The results are shown in Fig.~\ref{fig_co10_lf}. Compared to Fig.~\ref{fig_co_lf}, we have removed the $\langle z \rangle = 1.43$ bin from the 1mm data as the CO(2-1) line at 3mm is observed in practically the same redshift range and is subject to smaller uncertainties related to CO excitation corrections. Our observations succeed in sampling the predicted knee of the CO(1-0) luminosity functions at least up to $z\sim3$. Our measurements reveal that the knee of the CO(1-0) luminosity function shifts toward higher luminosities as we move from $z\approx 0$ \citep{keres03,boselli14} to $z\sim 2$. Our results agree with the model predictions at $z<1$. However, at $z>1$ they suggest an excess of CO--luminous sources, compared to the current models. 
This result is robust against CO excitation uncertainties: For example, it is already apparent in the $\langle z\rangle =1.43$ bin, where we covered the CO(2-1) line in our 3mm cube; this line is typically close to be thermalized in star forming galaxies, so excitation corrections are small. Our result is also broadly consistent with the findings by \citet{keating16}, based on a CO(1-0) intensity mapping study at $z=2-3$, that is unaffected by CO excitation.

\subsection{Cosmic H$_2$ mass density}

To derive H$_2$ masses, and the evolution of the cosmic H$_2$ mass density, we now convert the CO(1-0) luminosities into molecular gas masses $M_{\rm H2}$:
\begin{equation}\label{eq_H2}
M_{\rm H2}=\alpha_{\rm CO} \, L_{\rm CO(1-0)}'
\end{equation}
The conversion factor $\alpha_{\rm CO}$ implicitly assumes that CO is optically thick. The value of $\alpha_{\rm CO}$ critically depends on the metallicity of the interstellar medium \citep[see][for a review]{bolatto13}. A galactic value $\alpha_{\rm CO} = 3-6$ \Msun\,(\Kkmspc)$^{-1}$ is expected for most of non-starbursting galaxies with metallicities $Z\gsim 0.5$ Z$_\odot$ \citep{wolfire10,glover11,feldmann12}. At $z\sim 0.1$, this is the case for the majority of main-sequence galaxies with $M_{*}>10^9$\,\Msun{} \citep{tremonti04}. This seems to hold even at $z\sim 3$, if one takes into account the SFR dependence of the mass-metallicity relation \citep{mannucci10}. Following \citet{daddi10a}, we thus assume $\alpha_{\rm CO} = 3.6$ \Msun\,(\Kkmspc)$^{-1}$ for all the sources in our sample. In section \ref{sec_discussion} we discuss how our results would be affected by relaxing this assumption.

\begin{figure*}
\begin{center}
\includegraphics[width=0.79\textwidth]{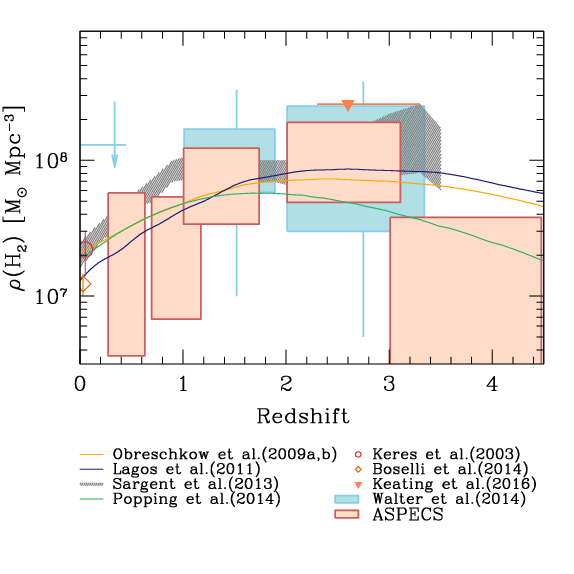}\\
\caption{Comoving cosmic mass density of molecular gas in galaxies $\rho$(H$_2$) as a function of redshift, based on our molecular survey in the UDF. Our ASPECS constraints are displayed as red boxes. The vertical size indicates our uncertainties (see text for details). Our measurements are not extrapolated to account for the faint end of the molecular gas mass function. Since our observations sample the expected knee of the CO luminosity functions in the redshift bins of interest, the correction is expected to be small ($<2\times$). Semi-analytical model predictions by \citet{obreschkow09a,obreschkow09b}, \citet{lagos12} and \citet{popping14a,popping14b} are shown as lines; the empirical predictions by \citet{sargent14} are plotted as a grey area; the constraints by \citet{keating16} are displayed with triangles; the PdBI constraints \citep{walter14} are represented by cyan boxes. Our ALMA observations show an evolution in the cosmic density of molecular gas up to $z\sim 4.5$. The global molecular content of galaxies at the peak of galaxy formation appears 3--10$\times$ higher than in galaxies in the local universe, although large uncertainties remain due to the limited area that is covered. 
}
\label{fig_rhoH2_z}\end{center}
\end{figure*}

Next, we compute the cosmic density of molecular gas in galaxies, $\rho$(H$_2$):
\begin{equation}\label{eq_rhoH2}
\rho ({\rm H_2})=\frac{1}{V}\,\sum_{i} \sum_{j=1}^{N_i}\frac{M_{i,j} \, P_j}{C_j}
\end{equation}
where $M_{i,j}$ is a compact notation for $M_{\rm H2}$ of the $j$-th galaxy in the mass bin $i$, and the index $i$ cycles over all the mass bins. As for $\Phi$, the uncertainties on $\rho$(H$_2$) are dominated by the Poissonian errors. Our findings are shown in Fig.~\ref{fig_rhoH2_z} and are summarized in Tab.~\ref{tab_rhoH2}. We note that the measurements presented here are only based on the observed part of the luminosity function. Therefore, we do not attempt to correct for undetected galaxies in lower luminosity bins given the large uncertainties in the individual luminosity bins and the unknown intrinsic shape of the CO luminosity function.

From Fig.~\ref{fig_rhoH2_z}, it is clear that there is an evolution in the molecular gas content of galaxies with redshift, in particular compared with the $z=0$ measurements by \citet{keres03} [$\rho$(H$_2$)=$(2.2\pm0.8)\times 10^7$\,\Msun{}\,Mpc$^{-3}$] and \citet{boselli14} [$\rho$(H$_2$)$= (1.2\pm0.2)\times 10^7$\,\Msun{}\,Mpc$^{-3}$]. The global amount of molecular gas stored in galaxies at the peak epoch of galaxy assembly is 3--10 times larger than at the present day. This evolution can be followed up to $z\sim4.5$, i.e., 90\% of the age of the universe. This trend agrees with the initial findings using PdBI \citep{walter14}. Our results are consistent with the constraints on $\rho$(H$_2$) at $z\sim 2.6$ based on the CO(1-0) intensity mapping experiment by \citet{keating16}\footnote{For a CO intensity mapping experiment based on the ASPECS data, see \citet{carilli16}.}: By assuming a linear relation between the CO luminosity of galaxies and their dark matter halo mass, they interpret their CO power spectrum constraint in terms of $\rho$(H$_2$) $<2.6\times10^{8}$\,\Msun{}\,Mpc$^{-1}$ (at 1-$\sigma$). They further tighten the constraint on $\rho$(H$_2$) by assuming that the $L_{\rm CO}$--dark matter halo mass relation has a scatter of 0.37 dex (a factor $\approx 2.3$), which translates into $\rho$(H$_2$)=$1.1_{-0.4}^{+0.7}\times10^8$\,\Msun{}\,Mpc$^{-1}$, in excellent agreement with our measurement. Our findings are also consistent with the global increase of the gas fraction as a function of redshift found in targeted observations \citep[e.g.,][]{daddi10a,riechers10,tacconi10,tacconi13,genzel10,genzel15,geach11,magdis12,magnelli12}, although we find a large variety in the gas fraction in individual sources \citep[see][]{decarli16}. Our results are also in general agreement with the expectations from semi-analytical models \citep{obreschkow09a,obreschkow09b,lagos11,lagos12,popping14a,popping14b} and from empirical predictions \citep{sargent12,sargent14}. From the present data, there is an indication for a decrease of $\rho$(H$_2$) at $z>3$, as suggested by some models\footnote{The $\rho$(H$_2$) value at $z>3$ in the models by \citet{popping16} is lower than in the predictions in \citet{lagos11}. This might be surprising as the CO(1-0) luminosity function in the former exceeds the one of the latter, especially at high redshift (see Fig.~\ref{fig_co10_lf}). This discrepancy is explained with the non-trivial galaxy--to--galaxy variations of $\alpha_{\rm CO}$ in the two models.}. A larger sample of $z>3$ CO emitters with spectroscopically--confirmed redshifts, and covering more cosmic volume, is required in order to explore this redshift range.

\begin{table*}
\caption{\rm Redshift ranges covered in the molecular line scans, the corresponding comoving volume, the number of galaxies in each bin (accounting for different line identifications), and our constraints on the molecular gas content in galaxies $\rho$(H$_2$) and $\rho$(ISM).} \label{tab_rhoH2}
\begin{center}
\begin{tabular}{cccccccccccc}
\hline
Transition  & $\nu_0$ & $z_{\rm min}$ & $z_{\rm max}$ & $\langle z \rangle$ & Volume & N(H$_2$) & log $\rho_{\rm min}$(H$_2$) & log $\rho_{\rm max}$(H$_2$) & N(ISM) & log $\rho_{\rm min}$(ISM) & log $\rho_{\rm max}$(ISM) \\
            & [GHz]   &     &        &       & [Mpc$^3$] &       & [\Msun{}\,Mpc$^{-3}$] & [\Msun{}\,Mpc$^{-3}$] &       & [\Msun{}\,Mpc$^{-3}$] & [\Msun{}\,Mpc$^{-3}$] \\
    (1)     & (2)     & (3) & (4)    & (5)   & (6)       & (7)   & (8)  & (9)  & (10) &(11)& (12)   \\
\hline
\multicolumn{12}{c}{1mm (212.032--272.001 GHz)}\\
CO(3-2) 	     & 345.796 & 0.2713  & 0.6309 & 0.4858 &  314 & 1--2 & 6.56 & 7.76 &  2 & 6.36 & 7.18 \\
CO(4-3) 	     & 461.041 & 0.6950  & 1.1744 & 0.9543 & 1028 & 0--5 & 6.83 & 7.73 &  5 & 7.13 & 7.60 \\
\hline
\multicolumn{12}{c}{3mm (84.176--114.928 GHz)}\\
CO(2-1) 	     & 230.538 & 1.0059  & 1.7387 & 1.4277 & 1920 & 3    & 7.53 & 8.09 & 13 & 7.50 & 7.77 \\
CO(3-2) 	     & 345.796 & 2.0088  & 3.1080 & 2.6129 & 3363 & 2--7 & 7.69 & 8.28 &  6 & 7.04 & 7.46 \\
CO(4-3) 	     & 461.041 & 3.0115  & 4.4771 & 3.8030 & 4149 & 0--5 & 5.53 & 7.58 &  0 & --   & 6.21 \\
\hline
\end{tabular}
\end{center}
\end{table*}

\subsection{Estimates from dust continuum emission}

\begin{figure}
\includegraphics[width=0.99\columnwidth]{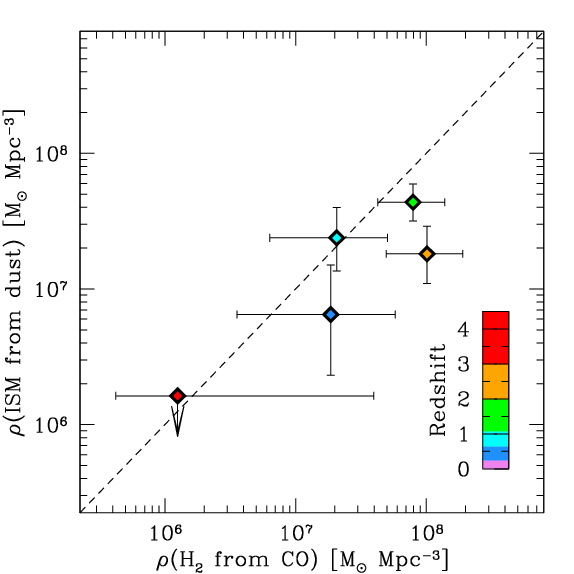}\\
\caption{Comparison between the CO-derived estimates of $\rho$(H$_2$) and the 1mm dust continuum-based estimates of $\rho$(ISM). The galaxies are binned in the same redshift bins as presented in Fig.~\ref{fig_rhoH2_z}, as indicated by the color of the symbols. The one--to--one case is shown as a dashed line. The dust--based estimates agree with the CO-based estimates at $z<2$, but they seem to fall below line of unity case at higher redshifts.}
\label{fig_rhoH2_co_ism}
\end{figure}

In Fig.~\ref{fig_rhoH2_co_ism} we compare the constraints on $\rho$(H$_2$) inferred from CO with those on $\rho$(ISM) derived from the dust continuum in our observations of the UDF. These are derived following \citet{scoville14}. In brief, for each 1mm continuum source (see the companion paper \citealt{aravena16a}), the ISM mass is computed as:
\begin{equation}\label{eq_scoville}
\frac{M_{\rm ISM}}{10^{10}\,{\rm M_\odot}}=\frac{1.78}{(1+z)^{4.8}}\,\,\frac{S_\nu}{\rm mJy} \,  \left(\frac{\nu}{\rm 350\,GHz}\right)^{-3.8} \, \frac{\Gamma_0}{\Gamma_{\rm RJ}} \, \left(\frac{D_{\rm L}}{\rm Gpc}\right)^2
\end{equation}
where $S_\nu$ is the observed continuum flux density, $\nu$ is the observing frequency (here, we adopt $\nu=242$\,GHz as the central frequency of the continuum image), $\Gamma_{\rm RJ}$ is a unitless correction factor that accounts for the deviation from the $\nu^2$ scaling of the Rayleigh-Jeans tail, $\Gamma_0=0.71$ is the tuning value obtained at low-$z$, and $D_{\rm L}$ is the luminosity distance \citep[see eq.~12 in][]{scoville14}. The dust temperature (implicit in the definition of $\Gamma_{\rm RJ}$), is set to 25\,K. The ISM masses obtained via eq.~\ref{eq_scoville} for each galaxy detected in the continuum \citep[see][]{aravena16a} are then split in the same redshift bins used for the CO-based estimates, and summed. We include here all the sources detected down to S/N=3 in the 1mm continuum. Poissonian uncertainties are found again to dominate the estimates of $\rho$ (if model uncertainties are neglected). The values of $\rho$(ISM) obtained in this way are reported in Tab.~\ref{tab_rhoH2}. We find that the ISM mass density estimates are roughly consistent (within the admittedly large uncertainties) with the CO-based estimates in the lower redshift bins ($z\sim 0.5$, $0.95$, and $1.4$), while discrepancies are found at $z>2$, where $\rho$(H$_2$) estimates based on CO tend to be larger than $\rho$(ISM) estimates based on dust. \citet{scoville15} present a different calibration of the recipe that would shift the dust-based mass estimates up by a factor $1.5$. However, even applying the more recent calibration would not be sufficient to significantly mitigate the discrepancy between CO-based and dust-based estimates of the gas mass at high redshift. In \citet{aravena16a} we show that all of our 1mm continuum sources detected at $>$3.5-$\sigma$ (except one) are at $z<2$. On the other hand, the redshift distribution of CO-detected galaxies in our sample extends well beyond $z$=2, thus leading to the discrepancy in the $\rho$ estimates at high redshift. Possible explanations for this difference might be related to the dust temperature and opacity, and to the adopted $\alpha_{\rm CO}$. A higher dust temperature in high-$z$ galaxies ($>40$\,K) would shift the dust emission towards higher frequencies, thus explaining the comparably lower dust emission observed at 1mm (at a fixed IR luminosity). Moreover, at $z=4$ our 1mm continuum observations sample the rest-frame $\sim 250$\,$\mu$m range, where dust might turn optically thick (thus leading to underestimates of the dust emission). Finally, we might be over-estimating molecular gas masses at high $z$ if the $\alpha_{\rm CO}$ factor is typically closer to the ULIRG/starburst value [$\alpha_{\rm CO}\approx 0.8$\,\Msun{}(\Kkmspc)$^{-1}$, see \citealt{daddi10b,bolatto13}]. However, the observed low CO excitation and faint IR luminosity do not support the ULIRG scenario for our high-$z$ galaxies. Furthermore, any metallicity evolution would yield a higher $\alpha_{\rm CO}$ at high $z$, instead of a lower one. In Paper IV we discuss the discrepancy between dust- and CO-based gas masses on a source-by-source basis.

\section{Summary and Discussion}\label{sec_discussion}

In this paper we use our ALMA molecular scans of the {\em Hubble} UDF in band~3 and band~6 to place blind constraints on the CO luminosity function up to $z\sim 4.5$. We provide constraints on the evolution of the cosmic molecular gas density as a function of redshift. This study is based on galaxies that have been blindly selected through their CO emission, and not through any other multi--wavelength property. The CO number counts have been corrected for by two parameters, {\em fidelity} and {\em completeness}, which take into account the number of false positive detections due to noise peaks and the fraction of lines that our algorithm successfully recovers in our data cubes from a parent population of known (artificial) lines.  

We start by constructing CO luminosity functions for the respective rotational transitions of CO for both the 3\,mm and 1\,mm observations. We compare these measurements to models that also predict CO luminosities in various rotational transitions, i.e. no assumptions were made in comparing our measurements to the models. This comparison shows that our derived CO luminosity functions lie above the predictions in the 3\,mm band. On the other hand, in the 1\,mm band our measurements are comparable to the models. Together this implies that the observed galaxies are more gas--rich than currently attributed for in the models, but with lower excitation.

Accounting for a CO excitation characteristic of main--sequence galaxies at $z\sim 1$--2, we derive the CO luminosity function of the ground--transition of CO (J=1--0) from our observations. We do so only up the J=4 transition of CO, to ensure that our results are not too strongly affected by the excitation corrections that would dominate the analysis at higher J. We find an evolution in the CO(1-0) luminosity function compared with observations in the local universe, with an excess of CO-emitting sources at the bright end of the luminosity functions. This is in general agreement with first constraints on the CO intensity mapping from the literature. This evolution exceeds what is predicted by the current models. This discrepancy appear to be a common trait of models of galaxy formation: galaxies with $M_*>10^{10}$\,\Msun{} at $z=2-3$ are predicted to be 2--3 times less star forming than observed (see, e.g., the recent review by \citealt{somerville15}), and similarly less gas--rich (see the analysis in \citealt{popping15a,popping15b}).

The sensitivity of the ALMA observations reaches below the knee of the predicted CO luminosity functions (around 5$\times$10$^{9}$\,\Kkmspc) at all redshifts. We convert our luminosity measurements into molecular gas masses via a  `Galactic' conversion factor. By summing the molecular gas masses obtained at each redshift, we obtain an estimate of the cosmic density of molecular gas in galaxies, $\rho$(H$_2$). Given the admittedly large uncertainties (mainly due to Poisson errors), and the unknown shape of the intrinsic CO luminosity functions, we do not extrapolate our measurements outside the range of CO luminosities (i.e., H$_2$ masses) covered in our survey.

We find an increase (factor of 3--10) of the cosmic density of molecular gas from $z\sim 0$ to $z\sim$ 2--3, albeit with large uncertainties given the limited statistics. This is consistent with previous findings that the gas mass fraction increases with redshift \citep[see, e.g.,][]{tacconi10,tacconi13,magdis12}. However our measurements have been derived in a completely different fashion, by simply counting the molecular gas that is present in a given cosmic volume, without any prior knowledge of the general galaxy population in the field. In this respect, our constraints on $\rho$(H$_2$) are actually lower limits, in the sense that they do not recover the full extent of the luminosity function. However, a) we do sample the predicted knee of the luminosity function in most of the redshift bins, suggesting that we recover a large part ($>$50\%) of the total CO luminosity per comoving volume; b) the fraction of the CO luminosity function missed because of our sensitivity cut is likely larger at higher redshift, i.e., correcting for the contribution of the faint end would make the evolution in $\rho$(H$_2$) even steeper.

We have also derived the molecular gas densities using the dust emission as a tracer for the molecular gas, following \citet{scoville14,scoville15}. The molecular gas densities derived from dust emission are generally smaller than but broadly consistent with those measured from CO at $z<2$, but that they might fall short at reproducing the predicted gas mass content of galaxies at $z>2$. 

Our analysis demonstrates that CO-based gas mass estimates result in 3--10 times higher gas masses in galaxies at $z\sim2$ than in the local universe. The history of cosmic SFR \citep{madau14} appears to at least partially follow the evolution in molecular gas supply in galaxies. The remaining difference between the evolution of the SFR density (a factor of $\sim 20$) and the one of molecular gas (a factor of 3--10) may due to the shortened depletion time scales. A further contribution to this difference may be ascribed to cosmic variance. The UDF in general (and therefore also the region studied here) is found to be underdense at $z>3$ \citep[e.g., Fig.~14 in][]{beckwith06} and in IR-bright sources \citep{weiss09}. 
The impact of cosmic variance can be estimated empirically from the comparison with the number counts of sources detected in the dust continuum \citep{aravena16a}, or analytically from the variance in the dark matter structures, coupled with the clustering bias of a given galaxy population \citep[see, e.g.,][]{somerville04}. \citet{trenti08} provide estimates of the cosmic variance as a function of field size, halo occupation fraction, survey completeness, and number of sources in a sample. For a $\Delta z=1$ bin centered at $z=2.5$, a 100\% halo occupation fraction and 5 sources detected over 1 arcmin$^2$ (i.e., roughly mimicing the $z\sim2.5$ bin in our analysis), the fractional uncertainty in the number counts due to cosmic variance is $\sim 20$\% ($\sim 60$\% if we include Poissonian fluctuations). 
Already a factor 5 increase in target area (resulting in a field that is approximately the size of the {\em Hubble} eXtremely Deep Field, \citealt{illingworth13}), at similar depth, would beat down the uncertainties significantly ($\lsim 30$\,\%, including Poissonian fluctuations). With ALMA now being fully operational, such an increase in areal coverage appears to be within reach.

\acknowledgements

We thank the anonymous referee for excellent feedback that improved the quality of the paper.
FW, IRS, and RJI acknowledge support through ERC grants COSMIC--DAWN, DUSTYGAL, and COSMICISM, respectively. M.A. acknowledges partial support from FONDECYT through grant 1140099. DR acknowledges support from the National Science Foundation under grant number AST-1614213 to Cornell University.  FEB and LI acknowledge Conicyt grants Basal-CATA PFB--06/2007 and Anilo ACT1417. FEB also acknowledge support from FONDECYT Regular 1141218 (FEB), and the Ministry of Economy, Development, and Tourism's Millennium Science Initiative through grant IC120009, awarded to The Millennium Institute of Astrophysics, MAS. IRS also acknowledges support from STFC (ST/L00075X/1) and a Royal Society / Wolfson Merit award. Support for RD and BM was provided by the DFG priority program 1573 `The physics of the interstellar medium'.  AK and FB acknowledge support by the Collaborative Research Council 956, sub-project A1, funded by the Deutsche Forschungsgemeinschaft (DFG). PI acknowledges Conict grants Basal-CATA PFB--06/2007 and Anilo ACT1417. RJA was supported by FONDECYT grant number 1151408. 
This paper makes use of the following ALMA data: \dataset[ ADS/JAO.ALMA\# 2013.1.00146.S and 2013.1.00718.S.]{https://almascience.nrao.edu/aq/}. ALMA is a partnership of ESO (representing its member states), NSF (USA) and NINS (Japan), together with NRC (Canada), NSC and ASIAA (Taiwan), and KASI (Republic of Korea), in cooperation with the Republic of Chile. The Joint ALMA Observatory is operated by ESO, AUI/NRAO and NAOJ. The 3mm-part of the ASPECS project had been supported by the German ARC.

\label{lastpage}

\end{document}